\documentclass[12pt]{iopart}
\usepackage{cite}
\usepackage{graphicx}
\usepackage{iopams}

\renewcommand{\eref}[1]{equation~(\ref{#1})}

\begin{document}

\title[Effective magnetic fields in atomic gases]{Effective magnetic fields
induced by EIT in ultra-cold atomic gases}

\author{G. Juzeli\={u}nas\dag, J. Ruseckas\dag\ and P. \"{O}hberg\ddag}

\address{\dag\ Institute of Theoretical Physics and Astronomy of Vilnius
University, A. Go\v{s}tauto 12, 01108 Vilnius, Lithuania}

\address{\ddag\ Department of Physics, University of Strathclyde, Glasgow G4
0NG, Scotland}

\ead{gj@itpa.lt}

\begin{abstract}
We study the influence of two resonant laser beams (to be referred to as the
control and probe beams) on the centre of mass motion of ultra-cold atoms
characterised by three energy levels of the $\Lambda$-type.  The laser beams
being in the Electromagnetically Induced Transparency (EIT) configuration drive
the atoms to their dark states. We impose the adiabatic approximation and obtain
an effective equation of motion for the dark state atoms.  The equation contains
a vector potential type interaction as well as an effective trapping potential.
We concentrate on the situation where the control and probe beams are
co-propagating and have Orbital Angular Momenta (OAM). The effective magnetic
field is then oriented along the propagation direction of the control and probe
beams.  Its spatial profile can be shaped by choosing proper laser beams.  We
analyse several situations where the effective magnetic field exhibits a radial
dependence. In particular we study effective magnetic fields induced by Bessel
beams, and demonstrate how to generate a constant effective magnetic field for a
ring geometry of the atomic trap.  We also discuss a possibility to create an
effective field of a magnetic monopole.
\end{abstract}


\submitto{\jpb}

\maketitle

\section{Introduction}

Recent experimental advances in trapping and cooling atoms have made it possible
to produce atomic Bose-Einstein Condensates (BECs)
\cite{ketterle95,hulet95,dalfovo99,bec_stri} and degenerate Fermi gases
\cite{demarco99,salomon01,ketterle03} at temperatures in the microkelvin range.
The atomic BECs and degenerate Fermi gases are the systems where an atomic
physicist often meets physical phenomena encountered in condensed matter
physics. For instance, atoms in optical lattices are often studied using the
Hubbard model \cite{dieter98} familiar from solid state physics.

Ultra-cold atomic gases have turned out to be a remarkably good medium for
studying a wide range of physical phenomena. This is mainly due to the fact that
it is relatively easy to experimentally manipulate parameters of the system,
such as the strength of interaction between the atoms, properties of a lattice
in which the atoms are trapped, the geometry of an external trap, etc.  Such a
freedom of manipulating parameters is usually not possible in other
systems known from condensed matter or solid state physics.

Atoms forming quantum gases are electrically neutral particles and there is no
vector potential type coupling of the atoms with a magnetic field. Therefore, a
direct analogy between the magnetic properties of degenerate atomic gases and
solid state phenomena is not necessarily straightforward. It is possible to
produce an effective magnetic field in a cloud of electrically neutral atoms by
rotating the system such that the vector potential will appear in the rotating
frame of reference \cite{bretin04,schweikhard04,baranov04}. This would
correspond to a situation where the atoms feel a uniform magnetic field. Yet
stirring an ultracold cloud of atoms in a controlled manner is a rather
demanding task.

There have also been suggestions to take advantage of a discrete periodic
structure of an optical lattice to introduce assymetric atomic transitions
between the lattice sites \cite{jaksch03,mueller04,sorensen04,osterloh05}. Using
this approach one can induce a nonvanishing phase for the atoms moving along a
closed path on the lattice, i.e.\ one can simulate a magnetic flux
\cite{jaksch03,mueller04,sorensen04,osterloh05}. However such a way of creating
the effective magnetic field is inapplicable to an atomic gas that does not
constitute a lattice.

A significant experimental advantage would be gained if a more direct way could
be used to induce an effective magnetic field. In previous papers
\cite{prl04,pra05}, we have shown how this can be done using two light beams in
an Electromagnetically Induced Transparency (EIT) configuration. Here we present
a more detailed analysis of the phenomenon for various spatial distributions of
the laser fields. We demonstrate that if at least one of these beams contains an
Orbital Angular Momentum (OAM), an effective magnetic field appears, which acts
on the electrically neutral atoms. In other words, the coupling between the
light and the atoms will provide an effective vector potential in the atomic
equations of motion. Compared to the rotating atomic gas, where only a constant
effective magnetic field is created \cite{bretin04,schweikhard04,baranov04},
using optical means will be advantageous since the effective magnetic field can
now be shaped by choosing proper control and probe beams. The appearance of the
effective vector potential is a manifestation of the Berry connection which is
encountered in many different areas of physics \cite{jack03,Sun90,Dum96}.

The outline of the paper is as follows.  In \sref{sec:form} we define a system
of three level atoms in the $\Lambda$-configuration interacting with the control
and probe beams.  We allow the two beams to have orbital angular momenta along
the propagation axis $z$. In \sref{sec:gen} we present a general treatment of
the adiabatic motion of multilevel atoms and apply it to derive the equation of
motion for the atom driven to the dark state by the control and probe beams of
light.  The resulting effective equation of motion contains the effective
trapping and vector potentials.  In sections \ref{sec:analysis} and
\ref{sec:spec} we analyse the effective magnetic field and effective trapping
potential in the case where at least one of the laser beams contains an orbital
angular momentum. We show that the spatial profile of the effective magnetic
field can be controlled by applying proper control and probe beams. We analyse
several situations where the effective magnetic field exhibits a radial
dependence. In particular we study effective magnetic fields induced by Bessel beams, 
and demonstrate how to generate a constant effective magnetic
field for ring geometry of the atomic trap. We also discuss a possibility to create 
an effective field of a magnetic monopole.
Finally in the
concluding \sref{sec:concl} we summarise the findings.

\section{\label{sec:form}Formulation}

\subsection{The atomic system}

\begin{figure}[tbp]
\begin{center}
\includegraphics[width=0.7\textwidth]{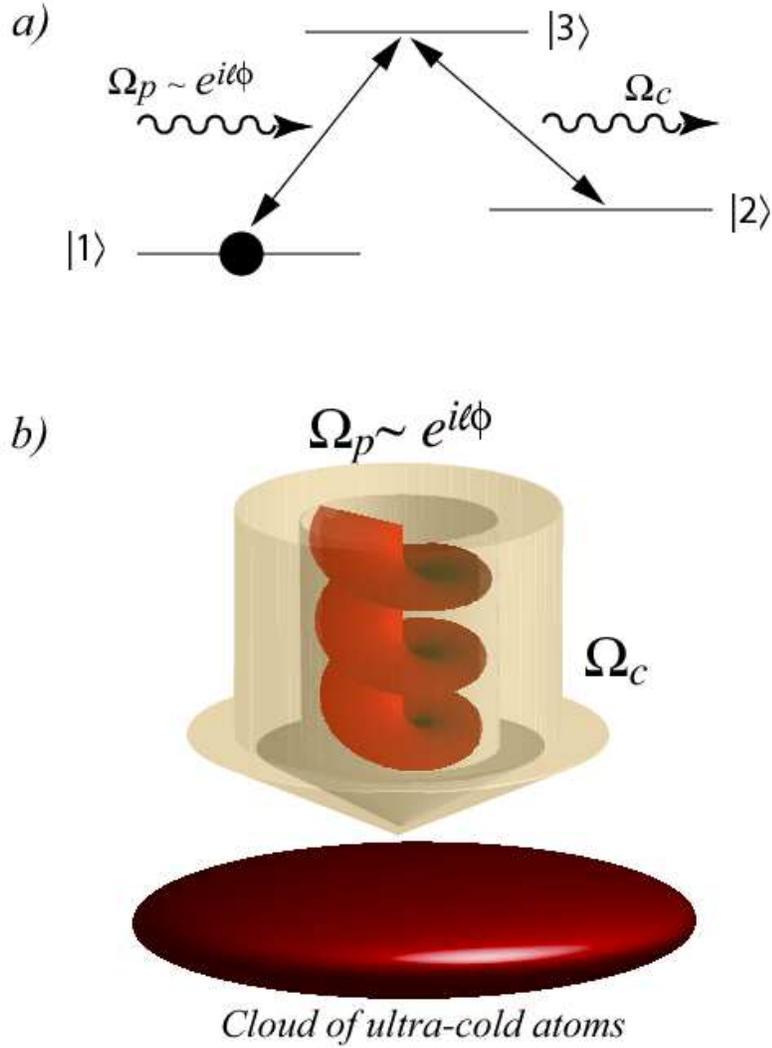}
\end{center}
\caption{a) The level scheme for the $\Lambda$ type atoms interacting with the
resonant probe beam $\Omega_p$ and control beam $ \Omega_c$. b) Schematic
representation of the experimental setup with the two light beams incident on
the cloud of atoms. The probe field is of the form
$\Omega_p\sim\rme^{\rmi\ell\protect\phi}$, where each probe photon carry an
orbital angular momentum $\hbar\ell$ along the propagation axis $ z$.}
\label{BildSchema}
\end{figure}

We shall consider an ensemble of atoms characterised by two hyper-fine ground
levels $1$ and $2$, as well as an electronic excited level $3$. The atoms
interact with two resonant laser beams in the EIT configuration (see
\fref{BildSchema}). The first beam (to be referred to as the control beam) drives
the transition $|2\rangle\rightarrow |3\rangle$, whereas the second beam (the
probe beam) is coupled with the transition $|1\rangle\rightarrow |3\rangle$, as
shown in \fref{BildSchema}a. The control laser has a frequency $\omega_c$,
a wave-vector $\bi{k}_c$, and a Rabi frequency $\Omega_c$. The probe field,
on the other hand, is characterised by a central frequency $\omega_p=ck_p$, a
wave-vector $\bi{k}_p$, and a Rabi frequency $\Omega_p$. Of special interest
is the case where the probe and control beams can carry OAM along the
propagation axis $z$. In that case, the spatial distribution of the beams is
given by \cite{allen99,oam}
\begin{equation}
\Omega_p=\Omega_p^{(0)}\rme^{\rmi(k_pz+l_p\phi)}
\label{omega-p}
\end{equation}
and
\begin{equation}
\Omega_c=\Omega_c^{(0)}\rme^{\rmi(k_cz+l_c\phi)},
\label{omega-c}
\end{equation}
where $\Omega_p^{(0)}$ and $\Omega_c^{(0)}$ are slowly varying amplitudes for
the probe and control fields, $\hbar\ell_p$ and $\hbar\ell_c$ are the
corresponding orbital angular momenta per photon along the propagation axis
$z$, and $\phi$ is the azimuthal angle.

\subsection{Hamiltonian for the electronic degrees of freedom of an atom}

Adopting the rotating wave approximation, the Hamiltonian for the electronic
degrees of freedom of an atom interacting with the control and probe fields is
in the rotating frame:
\begin{equation}
\hat{H}_0(\bi{r})=\epsilon_{21}|2\rangle\langle 2|+\epsilon
_{31}|3\rangle\langle 3|-\hbar(\Omega_p|3\rangle\langle 1|+\Omega
_c|3\rangle\langle 2|+\mathrm{H.c.})
\label{H-0}
\end{equation}
where $\epsilon_{21}=\hbar(\omega_2-\omega_1+\omega_c-\omega_p)$ and
$\epsilon_{31}=\hbar(\omega_3-\omega_1-\omega_p)$ are, respectively, the
energies of the detuning from the two- and single-photon resonances, with
$\hbar\omega_j$ being the electronic energy of the atomic level $j$. Note that
the spatial dependence of the Hamiltonian $\hat{H}_0(\bi{r})$ emerges
through the spatial dependence of the Rabi frequencies
$\Omega_p\equiv\Omega_p\left(\bi{r}\right)$ and $
\Omega_c\equiv\Omega_c\left(\bi{r}\right)$.

In what follows the control and probe fields are assumed to be
tuned to the two-photon resonance: $\epsilon_{21}=0$. The remaining two photon 
mismatch (if any) can be accommodated within the trapping potential
\begin{equation}
\hat{V}(\bi{r})=V_{1}(\bi{r})|1\rangle\langle 1|+ 
V_{2}(\bi{r})|2\rangle\langle 2|+
V_{3}(\bi{r})|3\rangle\langle 3|.
\label{V}
\end{equation}
where $V_j(\bi{r})$ is the trapping potential for an atom in the electronic
state $j$, with $j=1,2,3$.  Neglecting the two-photon detuning, the Hamiltonian
(\ref{H-0}) has the eigenstate
\begin{equation}
|D\rangle \equiv |D(\bi{r})\rangle =\frac{1}{\sqrt{1+|\zeta |^2}}
(|1\rangle -\zeta |2\rangle)
\label{D}
\end{equation}
characterised by the zero eigenenergy: $\hat{H}_0(\bi{r})|D\rangle=0$. Here
\begin{equation}
\zeta =\frac{\Omega_p}{\Omega_c}
\end{equation}
is the ratio of the amplitudes of the probe and control fields.

The state $|D\rangle$ is known as the dark state
\cite{Arimondo96,Harris97,eit,lukin03}. We shall be interested in a situation
where the atoms are driven to their dark states. If an atom is in the dark state
$|D\rangle$, the resonant control and probe beams induce the absorption paths
$|2\rangle\rightarrow |3\rangle$ and $|1\rangle\rightarrow |3\rangle$ which
interfere destructively, resulting in the Electromagnetically Induced
Transparency \cite {Arimondo96,Harris97,eit,lukin03}. In such a situation, the
transitions to the upper atomic level $3$ are suppressed, so the atomic level
$3$ is weakly populated. This justifies neglection of losses due to spontaneous
emission by excited atoms in the Hamiltonian (\ref{H-0}).

\section{\label{sec:gen}Equations of the atomic motion}

\subsection{Translational motion for a multilevel atom}

Let us now consider translational 
motion of an atom taking into account its internal degrees of freedom. 
The full atomic Hamiltonian is given by
\begin{equation}
\hat{H}=\frac{\hat{p}^2}{2M}+\hat{H}_0(\bi{r})+\hat{V}(\bi{r}),
\label{eq:full-H}
\end{equation}
where $\hat{\bi{p}}\equiv -\rmi\hbar\nabla$ is the momentum operator for the
atom positioned at $\bi{r}$, and $M$ is the atomic mass. The Hamiltonian for the
electronic degrees of freedom $\hat{H}_0(\bi{r})$ and the external trapping
potential $\hat{V}(\bi{r})$ featured in \eref{eq:full-H} are defined by
equations (\ref{H-0}) and (\ref{V}).

For fixed $\bi{r}$ the electronic Hamiltonian $\hat{H}_0(\bi{r})$ can be
diagonalised to give a set of eigenvectors $|X(\bi{r})\rangle$ and eigenvalues
$\varepsilon_X(\bi{r})$, where $X=D,+,-$.  The state with $X=D$ is the dark
atomic state given by \eref{D} and characterised by zero eigenenergy:
$\varepsilon_D(\bi{r})=0$. If the single and two photon detuning is zero, the
remaining eigenstates are:
\begin{equation}
|\pm(\bi{r})\rangle =\frac{1}{\sqrt{2}}(|B\rangle \pm |3\rangle),
\label{pm}
\end{equation}
corresponding eigenenergies being
\begin{equation}
\varepsilon_{\pm}(\bi{r})=\pm\Omega.
\label{e-pm}
\end{equation}
Here $\Omega = \sqrt{\Omega_p^2+\Omega_c^2}$ is the total Rabi frequency and
\begin{equation}
|B\rangle \equiv |B(\bi{r})\rangle =\frac{1}{\sqrt{1+|\zeta |^2}}
(\zeta^{*} |1\rangle +|2\rangle)
\label{B}
\end{equation}
is the atomic bright state.

The full atomic wave function $\Phi$ can then be expanded as:
\begin{equation}
|\Phi(\bi{r})\rangle =\sum_{X=D,+,-}\Psi_X(\bi{r})
|X(\bi{r})\rangle,
\label{eq:expansion}
\end{equation}
where a composite wavefunction $\Psi_X(\bi{r})$ describes the translational
motion of an atom in the electronic state $X$, with $X=D,+,-$.

Substituting \eref{eq:expansion} into the Schr{\"o}dinger equation
$\rmi\hbar\partial\Phi/\partial t =\hat{H}\Phi$, one arrives at a set of coupled
equations for the components $\Psi_X$. Introducing the column $\Psi
=\left(\Psi_D,\Psi_+,\Psi_-\right)^T$, it is convenient to represent these
equations in a matrix form:
\begin{equation}
\rmi\hbar\frac{\partial}{\partial t}\Psi =\left[\frac{1}{2M}(-\rmi\hbar\nabla -
\bi{A})^2+U\right]\Psi ,
\label{eq:matrix}
\end{equation}
where $\bi{A}$ and $U$ are the $3\times 3$ matrices with the following
elements:
\begin{eqnarray}
\bi{A}_{X,X'}=\rmi\hbar\langle X|\nabla X'\rangle,
\label{eq:A} \\
U_{X,X'}=\varepsilon_X(\bi{r})\delta_{X,X'}+\langle X|\hat{V}(
\bi{r})|X'\rangle,
\label{eq:V}
\end{eqnarray}
i.e.\ the matrix $U$ includes contributions from both the internal atomic
energies and also the external trapping potential.

Since the atomic internal motion is much faster than the external
(translational) one, the difference in the atomic energies $U_{X,X}-U_{X',X'}$
is normally much larger than the energies of non-adiabatic coupling between these
states. In such a situation, the translational motion of atoms in different
internal levels can be considered to be independent. This leads to the adiabatic
approximation.

\subsection{Effective equation of motion for a dark-state atom}

Specifically, let us suppose that the atomic dark state  $|D\rangle$ 
is well separated from the remaining 
atomic states $|\pm\rangle$. Neglecting
transitions to the latter states, \eref{eq:matrix} provides an
effective equation for the translational motion of an atom in the 
electronic dark state  $|D\rangle$:
\begin{equation}
\rmi\hbar\partial\Psi_D /\partial t=\hat{H}_{\mathrm{eff}}\Psi_D
\label{eq:matrix-q}
\end{equation}
where the effective Hamiltonian
\begin{equation}
\hat{H}_{\mathrm{eff}}=\frac{1}{2M}(-\rmi\hbar\nabla -
\bi{A}_{\mathrm{eff}})^2+V_{\mathrm{eff}}
\label{eq:H}
\end{equation}
is characterised by the effective vector and trapping potentials:
\begin{equation}
\bi{A}_{\mathrm{eff}}\equiv 
\bi{A}_{D,D}=\rmi\hbar\langle D|\nabla D\rangle
\label{A-D}
\end{equation}
and
\begin{equation}
V_{\mathrm{eff}}=U+\phi,
\label{V-eff}
\end{equation} 
with $U\equiv U_{D,D}$ being defined by the above \eref{eq:V}.  An additional
scalar potential $\phi$ appears due to the exclusion of the electronic states 
with $X=\pm$ in the effective equation of motion (\ref{eq:matrix-q}).  In particular, we have
\begin{eqnarray}
\phi & =\frac{1}{2M}\sum_{X=\pm}\bi{A}_{D,X}\bi{A}_{X,D}
\nonumber\\
& = \frac{\hbar^2}{2M}\left(\langle\nabla D|\nabla D\rangle
+\langle D|\nabla D\rangle\langle D|\nabla D\rangle\right)
\label{eq:fi1}
\end{eqnarray}
and
\begin{equation}
U = \frac{V_1(\bi{r})+|\zeta |^2 V_2(\bi{r})}{
1+|\zeta |^2} .
\label{U}
\end{equation} 
Since $V_1(\bi{r})$ and $V_2(\bi{r})$ are the trapping potentials for an atom in
the electronic states $1$ and $2$ respectively, $U$ represents the external
trapping potential for the atom in the dark state.

In this way, the full effective trapping potential $V_{\mathrm{eff}}$ is
composed of the external trapping potential $U$ and the geometric scalar
potential $\phi$.  The former $U$ is determined by the shape of the trapping
potentials $V_1(\bi{r})$ and $V_2(\bi{r})$, as well as the intensity
ratio $|\zeta|^2$. The latter geometric potential $\phi$ is determined
exclusively by the spatial dependence of the dark state $|D\rangle$ emerging
through the spatial dependence of the ratio between the Rabi frequencies
$\zeta=\Omega_p/\Omega_c$.  Note that the effective vector potential
$\bi{A}_{\mathrm{eff}}$ (known as a Berry connection \cite{jack03}) has a
geometric nature as well, because it also originates from the spatial dependence
of the atomic dark state $|D\rangle \equiv |D(\bi{r})\rangle$. 

\subsection{Adiabatic condition}

The energy difference between the dark state and the remaining atomic
states $|\pm(\bi{r})\rangle$ is characterised by the total Rabi frequency
$\Omega = \sqrt{\Omega_p^2+\Omega_c^2}$.  Assuming that the control and probe
fields are tuned to the one- and two-photon resonances
($\epsilon_{31},\epsilon_{21}\ll\hbar\Omega$), the adiabatic approach holds if
the non-diagonal matrix elements in \eref{eq:matrix} are much smaller than the
total Rabi frequency $\Omega$.  This leads to the following condition
\begin{equation}
F\ll\Omega
\label{condition-adiabatic}
\end{equation}
where the velocity-dependent term
\begin{equation}
F=\frac{1}{1+|\zeta |^2}\left|\nabla\zeta\cdot\bi{v}\right|
\label{F}
\end{equation}
reflects the two-photon Doppler detuning \cite{pra05}. Note that the condition
(\ref{condition-adiabatic}) does not accommodate effects due to the decay of the
excited atoms. The dissipation effects can be included replacing the energy of
the one-photon detuning $\epsilon_{31}$ by $\epsilon_{31}-\rmi\hbar\gamma_3$,
where $\gamma_3$ is the excited-state decay rate. In such a situation, the dark
state can be shown to acquire a finite lifetime
\begin{equation}
\tau_D\sim\gamma_3^{-1}\Omega^2/F^2
\label{Gamma-D}
\end{equation}
which should be large compared to other characteristic time scales of the
system.

The condition (\ref{condition-adiabatic}) implies that the inverse Rabi
frequency $\Omega^{-1}$ should be smaller than the time an atom travels a
characteristic length over which the amplitude or the phase of the ratio $\zeta
=\Omega_p/\Omega_c$ changes considerably. The latter length exceeds the optical
wavelength, and the Rabi frequency can be of the order of $ 10^7$ to $
10^8\,\mathrm{s}^{-1}$ \cite{hau99}. Therefore the adiabatic condition
(\ref{condition-adiabatic}) should hold for atomic velocities up to
tens of meters per second, i.e. up to extremely large velocities in the
context of ultra-cold atomic gases. The allowed atomic velocities become lower
if the spontaneous decay of the excited atoms is taken into account. 
The atomic dark state accquires then a finite lifetime $\tau_D$ equal to 
$\gamma_3^{-1}$ times the ratio $\Omega^2/F^2$, see Eq. (\ref{Gamma-D}). 
The atomic decay rate $\gamma_3$ is typically of the order $ 10^7\,\mathrm{s}^{-1}$.
Therefore if the atomic velocities are of the order of a
centimeter per second (a typical speed of sound in an atomic BEC), the atoms
should survive in their dark states up to a few seconds. This is comparable to
a typical lifetime of an atomic BEC.

\section{\label{sec:analysis}Analysis of the effective vector and trapping
potentials}

Substituting the expression (\ref{D}) for the dark state into 
\eref{A-D} for the effective vector potential, the latter takes the form:
\begin{equation}
\bi{A}_{\mathrm{eff}}=\rmi\hbar\frac{\zeta^*\nabla\zeta
-\zeta\nabla\zeta^*}{2(1+|\zeta|^2)}.
\label{eq:vector-2}
\end{equation}
The effective magnetic field reads then:
\begin{equation}
\bi{B}_{\mathrm{eff}}=\nabla\times\bi{A}_{\mathrm{eff}}
=\rmi\hbar\frac{\nabla\zeta^*\times\nabla\zeta}{(1+|\zeta|^2)^2}
\end{equation}
and the geometric scalar potential is
\begin{equation}
\phi=\frac{\hbar^2}{2M}\frac{\nabla\zeta^*\nabla\zeta}{(1+|\zeta|^2)^2}.
\end{equation}

\subsection{Separation into the amplitude and phase}

Let us express the ratio of Rabi frequencies $\zeta$ in terms of the amplitude
and phase:
\begin{equation}
\zeta =\frac{\Omega_p}{\Omega_c}=|\zeta |\rme^{\rmi S}.
\end{equation}
The effective vector potential, the effective magnetic field, and the effective
scalar potential then read
\begin{eqnarray}
\bi{A}_{\mathrm{eff}} = -\hbar\frac{|\zeta|^2}{1+|\zeta|^2}\nabla S, \\
\bi{B}_{\mathrm{eff}} = \hbar\frac{\nabla
S\times\nabla|\zeta|^2}{(1+|\zeta|^2)^2},\\
\phi = \frac{\hbar^2}{2M}\frac{(\nabla|\zeta|)^2+|\zeta|^2(\nabla
S)^2}{(1+|\zeta|^2)^2}.
\end{eqnarray}

\subsection{Representation in terms of the mixing angle}

It is convenient to introduce the mixing angle $\alpha$ via the following
relationships:
\begin{equation}
\sin\alpha =\frac{1}{\sqrt{1+|\zeta |^2}},\quad\cos\alpha =\frac{|\zeta
|}{\sqrt{1+|\zeta |^2}}.
\end{equation}
If the intensity ratio $|\zeta |^2$ is much larger than the unity, the mixing angle 
is $\alpha \approx 1/ \zeta $. On the other hand, if $|\zeta |^2 \ll 1$, 
we have $\alpha \approx \pi /2-|\zeta |$.

The dark state can now be represented as
\begin{equation}
|D\rangle =\sin\alpha |1\rangle -\cos\alpha\,\rme^{\rmi S}|2\rangle.
\end{equation}
The effective vector and scalar potentials can also be rewritten in terms of
the mixing angle:
\begin{equation}
\bi{A}_{\mathrm{eff}}=-\hbar\cos^2\alpha\nabla
S=-\frac{\hbar}{2}\left(1+\cos(2\alpha)\right)\nabla S
\label{eq:vector}
\end{equation}
and
\begin{eqnarray}
\phi &=\frac{\hbar^2}{2M}\left[\left(\frac{1}{2}\sin(2\alpha)\nabla
S\right)^2+(\nabla\alpha)^2\right]\nonumber\\
 & = \frac{\hbar^2}{8M}\left[\left(1-\cos^2(2\alpha)\right)(\nabla
S)^2+\frac{(\nabla\cos(2\alpha))^2}{1-\cos^2(2\alpha)}\right],
\label{eq:scalar}
\end{eqnarray}
i.e.\ both potentials can be expressed through the quantity
\begin{equation}
\cos(2\alpha)=\frac{|\zeta |^2-1}{|\zeta |^2+1}.
\label{cos-2aplha}
\end{equation}
The same applies to the effective magnetic field:
\begin{equation}
\bi{B}_{\mathrm{eff}}=\nabla\times\bi{A}_{\mathrm{eff}}
=\frac{\hbar}{2}\nabla S\times\nabla\cos(2\alpha).
\label{eq:B}
\end{equation}

\subsection{Co-propagating control and probe beams with OAM}

If the co-propagating probe and control fields carry OAM, their amplitudes
$\Omega_p$ and $\Omega_c$ are given by equations
(\ref{omega-p})--(\ref{omega-c}).  The phase of the ratio $\zeta
=\Omega_p/\Omega_c$ then reads
\begin{equation}
S=l\phi,
\label{S}
\end{equation}
where $l=l_p-l_c$. Note that although both the control and probe fields are
generally allowed to have non-zero OAM by equations
(\ref{omega-p})--(\ref{omega-c}), it is desirable for the OAM to be zero for one
of these beams. In fact, if both $l_p$ and $l_c$ were non-zero, the amplitudes
$\Omega_p$ and $\Omega_c$ should simultaneously go to zero along the $z$-axis.
In such a situation, the total Rabi frequency $\Omega
=\sqrt{\Omega_p^2+\Omega_c^2}$ would also vanish, leading to the violation of
the adiabatic condition (\ref{condition-adiabatic}) along the $z$-axis.

Substituting \eref{S} into equations (\ref{eq:vector}),
(\ref{eq:scalar}) and (\ref{eq:B}), one has
\begin{eqnarray}
\bi{A}_{\mathrm{eff}}=-\hbar\cos^2\alpha\frac{l}{\rho}\bi{e}_{
\varphi}, \label{eq:vector1} \\
\phi  = \frac{\hbar^2}{2M}\left[\left(\frac{1}{2}\sin(2\alpha)
\frac{l}{\rho}\right)^2+(\nabla\alpha)^2\right],
\label{eq:scalar1}
\end{eqnarray}
and
\begin{equation}
\bi{B}_{\mathrm{eff}}=\frac{\hbar}{2}\frac{l}{\rho}\bi{e}_{\varphi}
\times\nabla\cos(2\alpha)
\label{eq:B1}
\end{equation}
where $\bi{e}_{\varphi}$ is the unit vector in the azimuthal direction, and
$\rho$ is the cylindrical radius.

In what follows the intensity ratio $|\zeta |^2$ is considered to depend on the
cylindrical radius $\rho$ only (unless stated otherwise). In that case the effective scalar potential and
magnetic field reduce to
\begin{equation}
\phi =\frac{\hbar^2}{2M}\left[\left(\frac{1}{2}\sin(2\alpha)\frac{l}{
\rho}\right)^2+\left(\frac{\partial\alpha}{\partial\rho}\right)^2\right]
\label{eq:scalar2}
\end{equation}
and
\begin{equation}
\bi{B}_{\mathrm{eff}}=-\frac{\hbar}{2}\frac{l}{\rho}\frac{\partial}{
\partial\rho}\cos(2\alpha)\bi{e}_z .
\label{eq:B2}
\end{equation}
In such a situation the effective magnetic field is directed along the $z$-axis.

\section{\label{sec:spec}Specific cases}

\subsection{Polynomial case}

If we take
\begin{equation}
|\zeta | = a\rho +b\rho^2  ,
\end{equation}
then
\[
\cos(2\alpha)=\frac{(a\rho +b\rho^2)^2-1}{(a\rho +b\rho^2)^2+1}.
\]
Consequently one has:
\begin{eqnarray}
\bi{A}_{\mathrm{eff}} = -\hbar
l\frac{\rho(a+b\rho)^2}{1+\rho^2(a+b\rho)^2}\bi{e}_{\varphi},\\
\phi =\frac{\hbar^2}{2M}\frac{(l^2+1)a^2+2(l^2+2)ab\rho
+(l^2+4)b^2\rho^2}{(1+\rho^2(a+b\rho)^2)^2} .
\end{eqnarray}
In this case the effective magnetic field
\begin{equation}
\bi{B}_{\mathrm{eff}}=-\hbar l\frac{2(a+b\rho)(a+2b\rho)}{(1+\rho^2(a+b\rho
)^2)^2}\bi{e}_z
\end{equation}
exhibits a radial dependence.

\subsection{Bessel beam}

Suppose the probe field represents a Bessel beam and the Rabi frequency of the
control beam is almost constant within an atomic cloud. In such a case we have
\begin{equation}
\zeta =b J_l(a\rho)\rme^{\rmi l\varphi}
\end{equation}
where $b$ is a dimensionless constant determining the relative strength of the
probe field.  The effective vector and scalar potentials are then:
\begin{eqnarray}
\bi{A}_{\mathrm{eff}} = -\hbar\frac{b^2 J_l(a\rho)^2}{1+b^2 J_l(a\rho)^2}
\frac{l}{\rho}\bi{e}_{\varphi},\\
\phi = \frac{\hbar^2 b^2}{2M}\frac{4l^2J_l(a\rho)^2+
a^2\rho^2(J_{l-1}(a\rho)-J_{l+1}(a\rho))^2}{ 4\rho^2(1+b^2 J_l(a\rho)^2)^2} .
\end{eqnarray}
In this case the effective magnetic field
\begin{equation}
\bi{B}_{\mathrm{eff}}=-\hbar\frac{a b^2 l}{\rho}\frac{J_l(a\rho)
(J_{l-1}(a\rho)-J_{l+1}(a\rho))}{(1+b^2 J_l(a\rho)^2)^2}\bi{e}_z
\label{eq:B-bessel}
\end{equation}
also exhibits a radial dependence. Furthermore, the 
strength of the effective magnetic field alternates its sign, i.e. the
regions with the effective magnetic field aligned along $z$-axis are replaced by
the regions in which the effective magnetic field is directed opposite to the
$z$-axis and \textit{vice versa}.

Next we shall examine situations where the effective magnetic field is constant.

\subsection{Constant effective magnetic field for ring geometry}

In the previous paper \cite{pra05} we have analysed a constant effective field
in the case where the atomic motion is restricted to distances where
$\rho < \rho_{\mathrm{max}}$. This can be achieved if the 
intensity ratio is:
\begin{equation}
|\zeta |^2=\frac{\left(\rho /\rho_{\mathrm{max}}\right)^2}{1-\left(\rho
/\rho_{\mathrm{max}}\right)^2},
\label{zeta-exploding}
\end{equation}
so that the effective vector potential, \eref{eq:vector1}, takes the form
\begin{equation}
\bi{A}_{\mathrm{eff}}=-\hbar
l\rho\rho_{\mathrm{max}}^{-2}\bi{e}_{\phi}.
\label{A-constant}
\end{equation}
This yields a constant effective magnetic field
\begin{equation}
\bi{B}_{\mathrm{eff}}=-2\hbar l\rho_{\mathrm{max}}^{-2}\bi{e}_z ,
\label{B-constant}
\end{equation}
with the corresponding cyclotron frequency $\omega_c=\hbar 2l/M
\rho_{\mathrm{max}}^2$ and the magnetic length $\ell_B=\sqrt{\hbar /M\omega_c}
=\rho_{\mathrm{max}}/\sqrt{2l}$.

Let us now consider a situation where the atomic motion is restricted
additionally from below, i.e.\ $\rho >\rho_{\mathrm{min}}$.  In such a case the
constant effective magnetic field is obtained provided
\begin{equation}
|\zeta |^2=\frac{\rho^2-\rho_{\mathrm{min}}^2}{\rho_{\mathrm{max}}^2-\rho^2}.
\label{zeta-exploding1}
\end{equation}
The effective vector potential then takes the form
\begin{equation}
\bi{A}_{\mathrm{eff}}=-\hbar\frac{\rho^2-\rho_{\mathrm{min}}^2}{\rho_{
\mathrm{max}}^2 -\rho_{\mathrm{min}}^2}\frac{l}{\rho}\bi{e}_{\varphi} ,
\label{A-constant1}
\end{equation}
giving the following magnetic field strength:
\begin{equation}
\bi{B}_{\mathrm{eff}}=-\frac{2\hbar l}{\rho_{\mathrm{max}}^2
-\rho_{\mathrm{min}}^2}\bi{e}_z .
\label{B-constant1}
\end{equation}  
For $\rho\rightarrow \rho_{\mathrm{min}}$ and $\rho\rightarrow
\rho_{\mathrm{max}}$, the intensity ratio $|\zeta |^2$ goes  respectively to
zero and infinity, so the equations (\ref{zeta-exploding1})-(\ref{B-constant1})
have a meaning only within a disc where
$\rho_{\mathrm{min}}<\rho<\rho_{\mathrm{max}}$. In other words,
\eref{zeta-exploding1} can model an actual intensity distribution of the control
and probe beams only within this region. The effective magnetic flux over the
disc is given by $\Phi = 2\pi\hbar l$.  Since the winding number $l$ of light
beams can currently be as large as several hundreds \cite{courtial97, curtis02},
it is possible to induce a substantial flux $\Phi$ in the atomic cloud. This
might enable us to study phenomena related to filled Landau levels with a large
number of atoms in the quantum gas.

\begin{figure}[tbp]
\begin{center}
\includegraphics[width=0.7\textwidth]{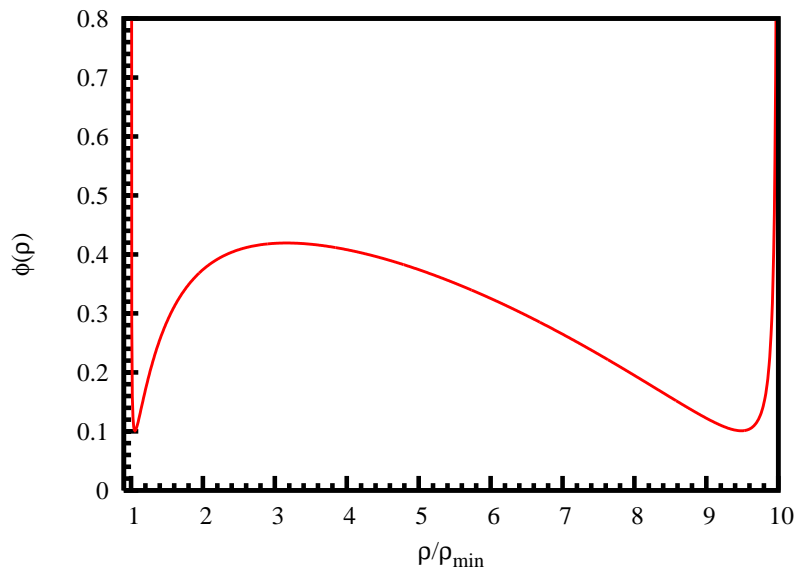}
\end{center}
\caption{Geometric scalar potential $\phi$ for the case of constant effective
magnetic field $B_{\mathrm{eff}}$ in the ring geometry.  The constants are
$M=1$, $\hbar=1$, $l=10$, $\rho_{\mathrm{min}}=1$, and
$\rho_{\mathrm{max}}=10$.}
\label{fig:ring}
\end{figure}

Finally the scalar potential is given by
\begin{equation}
\phi =\frac{\hbar^2}{2M}\left(\frac{l^2}{\rho^2}
\frac{(\rho_{\mathrm{max}}^2-\rho^2)(\rho^2-\rho_{\mathrm{min}}^2)}{
(\rho_{\mathrm{max}}^2-\rho_{\mathrm{min}}^2)^2}
+\frac{\rho^2}{(\rho_{\mathrm{max}}^2-\rho^2)(\rho^2-\rho_{\mathrm{min}}^2)}
\right).
\end{equation}
The potential $\phi$ has singuliarities both at $\rho =\rho_{\mathrm{min}}$ and
$\rho =\rho_{\mathrm{max}}$, as illustrated in \fref{fig:ring}. This might provide a 
natural trapping container confining the atoms within the ring.

\subsection{Effective magnetic field of a magnetic monopole}

The method for creating an effective magnetic field allows us to consider
various exotic cases. An interesting possibility is to create an effective field
of a magnetic monopole. Such an idea was proposed in \cite{Sun04}.

A possible expression for the vector potential of the monopole field has the
form
\[
\bi{A}\sim\frac{1-\cos\theta}{r\sin\theta}\bi{e}_{\varphi} ,
\]
where $r$, $\theta$ and $\varphi$ are the spherical coordinates.  Such a
potential has a singularity at the string where $\theta = 0$.  Unlike in
\cite{Sun04} we do not attempt to eliminate this singularity, since the
elimination does not make the field esier to create.

In order to get an effective vector potential of a magnetic monopole the
intensity ratio should obey the condition
\begin{equation}
|\zeta |^2=\frac{1-\cos\theta}{1+\cos\theta} .
\label{eq:zeta-monop}
\end{equation}
where we are no longer making the paraxial approximation.  In such a situation
\[
\cos(2\alpha)=-\cos\theta =-z/r .
\]
Consequently the effective vector potential, the effective magnetic field and
the scalar potential are
\begin{eqnarray}
\bi{A}_{\mathrm{eff}}=-\frac{\hbar l}{2}\frac{1-\cos\theta}{r\sin\theta}
\bi{e}_{\varphi},\\
\bi{B}_{\mathrm{eff}}=-\frac{\hbar l}{2r^2}\bi{e}_r ,
\label{eq:B-monop}\\
\phi =\frac{\hbar^2}{2M}\frac{l^2+1}{4r^2} .
\end{eqnarray}
The magnetic charge of the effective monopole is now proportional to the
difference of the orbital angular momentum of the light beams $\hbar l \equiv
\hbar(l_p-l_c)$. On the other hand, the emerging scalar potential $\phi$ is
repulsive and spherically symmetrical, and is characterised by the $r^{-2}$
behaviour. 

In order to satisfy the condition (\ref{eq:zeta-monop}) the Rabi frequencies
should obey the following:
\begin{eqnarray}
|\Omega_p|^2=f(\bi{r})(1-\cos\theta), \\
|\Omega_c|^2=f(\bi{r})(1+\cos\theta)
\end{eqnarray}
where $f(\bi{r})$ is an arbitrary function of the coordinates. For a light beam
with an OAM, its intensity is known to be zero along the propagation axis $z$
\cite{allen99,oam}. Therefore if the probe beam has an OAM, the function
$f(\bi{r})$ should be zero for $\cos\theta =-1$, i.e. along the negative part of
$z$-axis.  Under this condition the control beam should also be zero along the
negative part of $z$-axis, so the adiabatic condition
(\ref{condition-adiabatic}) is violated there.  Similar conclusions can be
reached if the control beam has OAM. 

In this way the effective field of a magnetic monopole cannot be created in the
whole space, i.e. the effective field differs from the field of a monopole in
the vicinity of the negative (or positive) part of the $z$-axis.  This
conclusion is valid even if the singularity of the potential is eliminated
\cite{Sun04}, since the intensities of the beams remain of the same form using
such a procedure.  It should be noted that a posibilitiy of creating the field
of a magnetic monopole can be improved applying a more complex scheme where
three laser beams act resonantly on four-level atoms in the tripod configuration
\cite{prl-nonabelian}.

\section{\label{sec:concl}Conclusions}

In this paper we have studied the effects of using probe and control beams with
orbital angular momentum in an EIT configuration. The nontrivial phase and
intensity of the incident light beams gives rise to effective magnetic vector
potentials and trapping potentials for the atoms. The theory holds for both
fermions and bosons where the effects from collisional interactions can be
readily included \cite{pra05}. Recent advances in spatial light modulator
technology enables us to consider rather exotic light beams \cite{mcgloin03}.
Indeed one of the advantages of using light to create the effective vector
potential, and consequently an effective magnetic field, is the freedom to
choose almost any spatially dependent effective magnetic field, as long as the
corresponding light fields obey Maxwell's equations. This means we have complete
freedom to choose the effective magnetic field in a two-dimensional geometry,
but we can also control the effective field in three dimensions.  Shaping light
beams in three dimensions is more difficult but certainly not impossible
\cite{whyte05}. We have analysed different cases where the radial dependence of
the magnetic fields was exploited. In particular the homogenous magnetic field
in a ring geometry and magnetic fields using Bessel beams were studied.

An effective magnetic field acting on an atomic quantum gas offers some truly
remarkable possibilities. We are now in a position to study magnetic effects
encountered in solid state situations with electrons. The effective magnetic
field can also be applied to investigate other intriguing phenomena which
intrinsically depend on the magnetic field.  For instance, the properties of a
gas described by a single completely filled Landau level can now be explored
using a cold gas of electrically neutral atomic fermions \cite{pra-hall}. In
addition, if the collisional interaction between the atoms is taken into
account, we can study the magnetic properties of a superfluid atomic Fermi gas
\cite{zwierlein03,regal04,salomon04,zwierlein05} where insight into the BCS-BEC
crossover regime could be gained.

\ack

This work was supported by the Royal Society of Edinburgh, the Alexander von
Humboldt Foundation and the Marie-Curie Trainings-site at the University of
Kaiserslautern. Helpful discussions with M. Fleischhauer are gratefully
acknowledged.

\end{document}